\documentclass[aps,prl,twocolumn,showpacs,superscriptaddress]{revtex4-1}
\usepackage{graphicx}
\usepackage{amsmath}
\usepackage{epstopdf}

\begin{document}

\title{Domain Wall Spin Dynamics in Kagome Antiferromagnets}

\author{E. Lhotel}
\email[]{elsa.lhotel@grenoble.cnrs.fr}
\affiliation{Institut N\'eel, CNRS \& Universit\'e Joseph Fourier, BP 166, 38042 Grenoble Cedex 9, France}
\author{V. Simonet} \affiliation{Institut N\'eel, CNRS \& Universit\'e Joseph Fourier, BP 166, 38042 Grenoble Cedex 9, France}
\author{J. Ortloff} \affiliation{Institut N\'eel, CNRS \& Universit\'e Joseph Fourier, BP 166, 38042 Grenoble Cedex 9, France}
 \affiliation{Institute for Theoretical Physics, University of W\"urzburg, Germany}
\author{B. Canals} \affiliation{Institut N\'eel, CNRS \& Universit\'e Joseph Fourier, BP 166, 38042 Grenoble Cedex 9, France}
\author{C. Paulsen} \affiliation{Institut N\'eel, CNRS \& Universit\'e Joseph Fourier, BP 166, 38042 Grenoble Cedex 9, France}
\author{E. Suard} \affiliation{Institut Laue Langevin, BP 156, 38042 Grenoble Cedex 9, France}
\author{T. Hansen} \affiliation{Institut Laue Langevin, BP 156, 38042 Grenoble Cedex 9, France}
\author{D. J. Price} \affiliation{Univ Glasgow, School of Chemistry, WestCHEM, Glasgow G12 8QQ, Scotland}
\author{P. T. Wood} \affiliation{Univ Cambridge, Chem Lab, Cambridge CB2 1EW, England}
\author{A. K. Powell} \affiliation{Institute of Inorganic Chemistry, Karlsruhe Institute of Technology, Engesserstrasse 15, D-76131 Karlsruhe, Germany.} \affiliation{Institute for Nanotechnologie, Karlsruhe Institute of Technology, Postfach 3640, D-76021 Karlsruhe, Germany }
\author{R. Ballou} \affiliation{Institut N\'eel, CNRS \& Universit\'e Joseph Fourier, BP 166, 38042 Grenoble Cedex 9, France}
\date{\today}

\begin{abstract}
We report magnetization and neutron scattering measurements down to 60 mK on a new family of Fe based kagome antiferromagnets, in which a strong local spin anisotropy combined with a low exchange path network connectivity lead to domain walls intersecting the kagome planes through strings of free spins. These produce unfamiliar slow spin dynamics in the ordered phase, evolving from exchange-released spin-flips towards a cooperative behavior on decreasing the temperature, probably due to the onset of long-range dipolar interaction. A domain structure of independent magnetic grains is obtained that could be generic to other frustrated magnets. 
\end{abstract}

\pacs{75.60.Ch, 75.40.Gb, 75.25.-j, 75.50.Ee}
\maketitle

Ordered magnets ordinarily display fragmentations into magnetic domains, interrelated to each other by the symmetries lost at the ordering. Such domains have mostly been studied in ferromagnetic materials \cite{ferro}. Investigations of antiferromagnetic domains are more elusive due to the absence of a spontaneous magnetization and to ultra-fast spin dynamics \cite{Kimel2004}. In all instances the domain walls might exhibit cooperative slow dynamics, but individual spins are never free. A paradigm for protected spins might emerge from topological frustration, which has provided an incredible reservoir in the search of novel magnetic phases \cite{Moessner2006,Lacroix}. An untackled question though is the influence of the lattice topology on the domain wall spin dynamics. We report here on the importance of the low connectivity of frustrated lattices such as the corner-sharing-triangle kagome one, which allows spins inside a domain-wall to be free from exchange interactions.

Topologically frustrated lattices may produce in extreme cases highly degenerate ground states, which inhibit magnetic ordering and lead to disordered phases with short-range spin-spin correlations and remarkable excitations \cite{Moessner2006,Lacroix}. In the classical Heisenberg kagome lattice with antiferromagnetic nearest-neighbor (NN) interactions, a strongly correlated paramagnetic state (spin liquid) with 120$^{\circ}$ spin arrangements on each triangle is expected down to the lowest temperature \cite{Ritchey1993}. Another example is the disordered spin ice ground state discovered in some pyrochlore materials, where two spins point into and two out of each corner-sharing tetrahedron \cite{Harris1997}. This was shown to result from exchange and dipolar interactions associated with a strong multiaxial anisotropy. Additional parameters in the Hamiltonian beyond the NN interactions  (next neighbors exchange interaction, single-ion anisotropy, dipolar interaction, Dzyaloshinskii-Moriya (DM) interaction, non-stoichiometry...) may also release the frustration usually leading to  complex magnetic orderings \cite{Lacroix, Matan}. In the kagome antiferromagnet for instance, a multiaxial anisotropy characterized by a three-fold direction of the spins in a triangle will gradually fix the spin orientation in the whole lattice. This lifts the massive ground state degeneracy leading to a non-collinear magnetic order. 

\begin{figure}
\includegraphics[keepaspectratio=true, width=9cm]{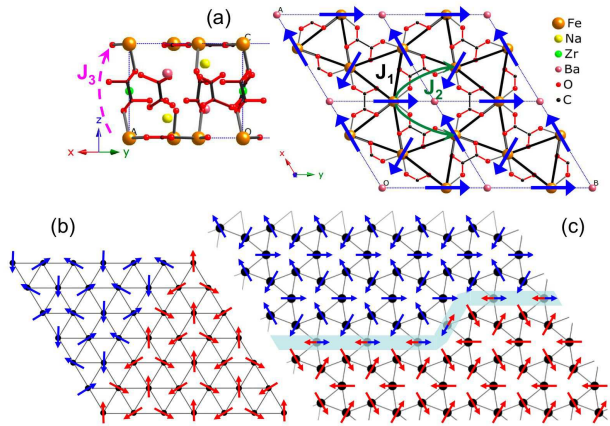}
\caption{(Color online) (a): Projected structure of QO-FeZr on the ($a$,$b$) plane (right) and the ($b$,$c$) plane (left) with $a=b=10.45$ \AA, $c=7.54$ \AA. There are three Fe$^{\rm II}$ per unit cell at positions (0, 0.6145, 0), (0.6145, 0, 0) and (0.3854, 0.3854, 0). The black lines materialize the Fe$^{\rm II}$ NN exchange interaction lattice. The NNN $J_2$ and $J_3$ exchange interactions are shown by the green and dashed pink arrows respectively. The blue arrows represent the ordered magnetic moments. The 180$^{\circ}$ antiferromagnetic domains (red and blue) are shown in the triangular lattice (b) and in the QO-FeA distorted kagome lattice (c), with a string of exchange-released spins along the domain wall.}
\label{figstructure}
\end{figure}

We have investigated such a model system with new metallo-organic compounds, built from Fe$^{\rm II}$ and bridged by C$_2$O$_4^{2-}$ oxalate ligands. Two new isostructural series were synthesized using hydrothermal methods: series I with the composition Na$_2$Ba$_3$[Fe$^{\rm II}_3$(C$_2$O$_4$)$_6$][A$^{\rm IV}$(C$_2$O$_4$)$_3$] where A$^{\rm IV}$ = Sn$^{\rm IV}$, Zr$^{\rm IV}$; and series II with the composition Na$_{2}$Ba$_3$[Fe$^{\rm II}_3$(C$_2$O$_4$)$_{6}$][A$^{\rm III}$(C$_2$O$_4$)$_3$]$_{0.5}$[A$^{\rm III}$(C$_2$O$_4$)$_2$(H$_2$O)$_2$]$_{0.5}$, where A$^{\rm III}$ = Fe$^{\rm III}$, Al$^{\rm III}$. In the following, these quinternary oxalate compounds will be abbreviated QO-FeA referring to the common divalent Fe$^{\rm II}$ and the cation A=Sn$^{\rm IV}$, Zr$^{\rm IV}$, Fe$^{\rm III}$ (QO-FeAl was not considered in the present study). 
They crystallize in the chiral trigonal P321 space group. The only magnetic ions are the Fe$^{\rm II}$ except in QO-FeFe where paramagnetic Fe$^{\rm III}$ in the low spin state (S=1/2) are present between the kagome planes, without significant interactions down to the lowest temperature. The magnetic Fe$^{\rm II}$ network forms, in the ($a$, $b$) plane, a distorted kagome lattice stacked along the $c$ axis, topologically equivalent to the kagome one if NN interactions only are considered (See Fig. \ref{figstructure}). All QO-FeA compounds present the same magnetic properties driven in particular by a strong single-ion anisotropy and weaker exchange interactions. These parameters explain the magnetic structure determined by neutron diffraction and the magnetization measurements, as briefly reported hereafter. Although the frustration is actually released by the anisotropy, the lattice topology maintains spin degrees of freedom associated with defects inherent to the magnetic structure. The signature of these quasi-Ising free spins in the ordered state is subsequently described in this letter.

We measured the magnetization and AC susceptibility of powder samples of the three compounds by the extraction method, using a purpose-built magnetometer and a Quantum Design MPMS magnetometer for temperatures above 2 K, and a superconducting quantum interference device magnetometer equipped with a miniature dilution refrigerator developed at the Institut N\'eel for temperatures down to 65 mK. Measurements were carried out for frequencies between 1.1 mHz and 5.7 kHz (more than six decades), with an applied AC field of 0.5 Oe. Powder neutron diffraction measurements were performed on the two two-axis diffractometers D20 and D2B with a wavelength equal to 2.4 \AA\ at the Institut Laue-Langevin high-flux reactor, Grenoble, France. Diffractograms were recorded down to 2~K on the three compounds (deuterated for series II), and down to 60~mK on the QO-FeZr compound. 

\begin{figure}
\includegraphics[keepaspectratio=true, width=7cm]{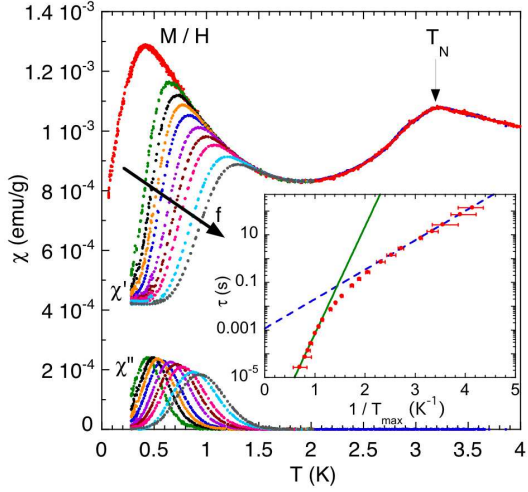}
\caption{(Color online) AC and DC susceptibility {\it vs} temperature: $M/H$ in an applied field $H_{DC}$=500 Oe (red circles), real part $\chi'$ and imaginary part $\chi"$ of the AC susceptibility with $H_{AC}$=1 Oe and 0.21 Hz$<f<$211 Hz. The inset shows $\tau$ vs $1/T_{max}$ in a semi-logarithmic plot. The error bars indicate the uncertainty in the determination of the $\chi''$ maximum. The lines are fits to the Arrhenius law with $\tau_{01}=2.1\times10^{-8}$ s and $E_1=$10 K (full line) and $\tau_{02}=1.1\times10^{-3}$ s and $E_2=$2.8 K (dashed line).} 
\label{figXT}
\end{figure}

\begin{figure}
\includegraphics[bb=10 20 308 190,scale=0.85]{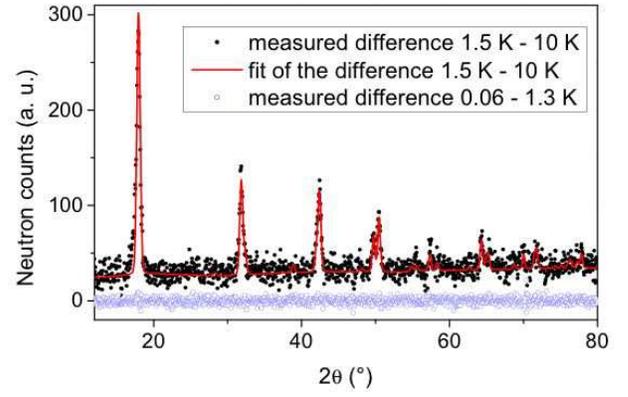}
\caption{(Color online) Magnetic diffraction pattern of QO-FeZr obtained from the difference between the diffractograms measured at 1.5 and 10 K on D2B. The red line is a fit with a propagation vector (0,0,1/2) and refined magnetic moment of 5.2(2) $\mu_B$. The absence of magnetic rearrangement is shown from the flat difference between the 0.06 and 1.3 K diffractograms. } 
\label{figneutrons}
\end{figure}

In the QO-FeA, the transition to an antiferromagnetic order is evidenced by a cusp in the magnetization at the N\'eel temperature $T_N=3.2$ K (See Fig. \ref{figXT}) and the rise below $T_N$ of magnetic Bragg peaks, as seen in powder neutron diffraction (See Fig. \ref{figneutrons}). The magnetic structure refinement indicates an antiferromagnetic stacking along the ${\bf c}$ axis and the so-called {\bf q=0} in-plane arrangement consisting of magnetic moments at 120$^{\circ}$ from each other and lying along the ${\bf a}$, ${\bf b}$ and $-{\bf a}-{\bf b}$ axes, with the same spin chirality for all the triangles (see Fig. \ref{figstructure}(a)) \cite{footnote1}. 

The energy scale of the main interactions can be estimated from the Curie-Weiss temperature $\theta$. The linear susceptibility $\chi=M/H$ was fitted in the range [50-300 K] using a Curie-Weiss model $C/(T-\theta)$ with $C={\cal N}_A\mu_{\rm eff}^2 / 3k_B$. This yields $\mu_{\rm eff}=6.4\ \mu_B$ ($S$=2, $L\approx$2) and $\theta=-5$ K, giving 3 K for the exchange energy $E^{NN}_{exch}$ due to the antiferromagnetic NN $J_1$ exchange interactions. This value is consistent with those reported in other compounds where the superexchange interactions are mediated by C$_2$O$_4^{2-}$ oxalate ligands \cite{Mennerich08, Kikkawa05,Nunez01}. $J_1$ is expected to be much stronger than the next nearest neighbor (NNN) $J_2$ (in-plane) and $J_3$ (inter-plane) interactions since the $J_2$ and $J_3$ exchange paths are longer and involve two C atoms (see Fig. \ref{figstructure}(a)) \cite{Price01,Fei05}. 

The low degree of magnetic frustration in the QO-FeA, estimated from $T_N/|\theta|\approx$ 1, is due to the large multiaxial magnetocrystalline anisotropy. The Fe octahedral symmetries can favor a moment orientation along the structural twofold axis, resulting in a different axis for each spin of the triangle at 120$^{\circ}$ from each other, as observed. The 3-dimensional ordering is ultimately stabilized via antiferromagnetic interplane interactions, that can be much weaker than T$_N$. An anisotropy energy of 10 K is inferred from the energy barrier determined by AC susceptibility in the single spin-flip regime as explained below, which agrees with the anisotropy reported in other Fe$^{\rm II}$ oxalate compounds \cite{Echigo08,Simizu88,Sledzinska86,Mennerich08}. 

\begin{figure}
\includegraphics[keepaspectratio=true, width=8.5cm]{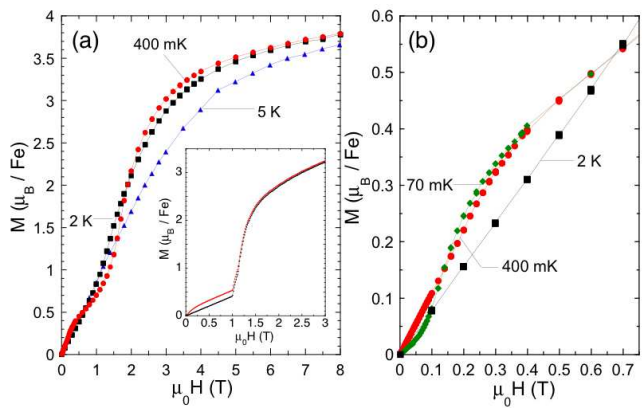}
\caption{(Color online) (a) Measured magnetization $M$ {\it vs} $H$ in QO-FeZr at 400 mK (red circles), 2 K (black squares) and 5 K (blue triangles). Inset: in black, calculated T = 0 K powder magnetization with $E^{NN}_{exch}$=3 K, $E^{inter}_{exch}$=0.3 K, $E_{anis}$=10 K and a magnetic cell doubled along ${\bf c}$. Calculations including dipolar interactions are not significantly different. In red,  same calculation but with the exact powder average contribution of 5\% of the paramagnetic quasi-Ising spins belonging to the domain walls computed at 400 mK. (b) Zoom of $M$ {\it vs} $H$ at low field: 2 K (black squares), 400 mK (red circles), 70 mK (green squares). } 
\label{figMH}
\end{figure}

The Hamiltonian of the QO-FeA was validated by comparing its exact ground states at zero Kelvin to measurements, as follows. The model including antiferromagnetic NN and inter-plane exchange interactions corresponding to energies $E^{NN}_{exch}$=3 K and $E^{inter}_{exch}$=0.3 K, and a multiaxial anisotropy term $E_{anis}$=10 K,  yields the observed magnetic structure. The powder averaged magnetization {\it vs} magnetic field, assuming a magnetic cell doubled along the $c$ axis, was also computed \cite{footnote2}. Some features in the magnetization curves below $T_N$ are reproduced: a metamagnetic process at $\approx$ 1 T and a non saturated magnetization at 8 T (see Fig. \ref{figMH}(a)). 

Whereas neutron diffraction in zero field has proven that there is no change of the magnetic structure itself down to 60 mK (See Fig. \ref{figneutrons}), additional features appear in the magnetization on lowering the temperature, associated with slow spin dynamics as probed by AC susceptibility. Below 2~K, there is a frequency dependence of the real part $\chi'$ and imaginary part $\chi''$ of the AC susceptibility (See Fig. \ref{figXT}). This is intrinsic to the system, since the same behavior was observed in several QO-FeZr samples from different batches as well as in QO-FeSn and QO-FeFe compounds. Measurements of magnetization relaxation {\it vs} time show that most of the magnetization goes to zero in a very short time ($<$ 10 s) at 65 mK. These observations prove that there is no strong pinning in the system and that only a small fraction of the quasi-Ising spins, estimated $\approx$ 5\%, is concerned with the dynamics. Assuming that the dynamics is governed by a single relaxation time $\tau$, $\chi"(T)$ is maximum when the measurement time ($=1/2 \pi f$) is equal to $\tau$.  In a usual thermal activated process over an energy barrier $E$, $\tau$ follows an Arrhenius law $\tau=\tau_0 \exp(E/k_BT)$, where $\tau_0$ is the characteristic relaxation time. Here, the plot of $\tau$ {\it vs.} $1/T_{max}$ (see inset of Figure \ref{figXT}) reflects the need to consider two distinct temperature regimes.

The "high" temperature regime ($T>0.8$ K) can be fitted by an Arrhenius law with ${\tau_0}_1 \approx 2 \times 10^{-8}$ s and $E_1=10$ K (full line in the inset of Fig. \ref{figXT}). This is consistent with single spin flips over the anisotropy barrier. That such spins can freely flip in a 3-dimensional ordered antiferromagnet is unusual. The key to this behavior resides in the influence of the lattice topology on the spins at the boundary between antiferromagnetic domains. As a result of the symmetry lowering at the phase transition, two 180$^{\circ}$ domains, where all the spins are reversed, coexist in the kagome planes. The single atomic distance width of the domain walls is caused by the strong anisotropy. Due to the low connectivity, a boundary spin is only shared by two triangles belonging to each domain and is blind to its neighbors along the domain wall \cite{footnote3}. The energy resulting from its interaction with the other spins is therefore the same in either of its two possible orientations and this spin is free to flip over the anisotropy barrier. This is illustrated in Fig. \ref{figstructure} where antiferromagnetic domains are schematized in the QO-FeA lattice (Fig. \ref{figstructure}(c)) and in a triangular lattice (Fig. \ref{figstructure}(b)) with a larger connectivity inhibiting the presence of the free spins. In the QO-FeA compounds, the single spin-flips are incoherent and should not result in a global motion of the string-like domain walls. The size of the antiferromagnetic domains could be roughly estimated from the neutron diffractograms. The broadening of the magnetic Bragg peaks with respect to the nuclear ones (resolution limited) yields, using the Sherrer equation \cite{Klug, Wan}, an average domain diameter of $\approx$500 \AA. This is in agreement with the domain size computed for 5\% of spins within the domain walls.

Below 0.8 K, the thermal single spin flip mechanism becomes too slow and a crossover is observed towards a more efficient process. The latter can be described below $\approx$ 500 mK by an Arrhenius law (dashed line in the inset of Fig. \ref{figXT}) with an abnormally high characteristic relaxation time ${\tau_0}_2 \approx 10^{-3}$~s and a reduced energy barrier $E_2=2.8$ K. The dipolar energy was estimated to be $\approx 0.2$ K, of the proper order of magnitude to explain this dynamical cross-over. Long-range dipolar interactions could start coupling the free spins along the domain walls whereas the anisotropy energy barrier may be partially erased by quantum tunneling. The full understanding of this dynamics is not achieved yet but it is interesting to note its similarity with what is reported in the pyrochlore spin ice materials \cite{Snyder2004,Jaubert09, Jaubert11}. The spin-ice dynamics is characterized by a high temperature regime of single spin-flips above an anisotropy barrier \cite{Matsuhira}. Then, below 10 K, quantum tunneling initiates another regime with a larger $\tau_0$, where spin-flips can be described as deconfined magnetic excitations called monopoles, which become frozen by dipolar interactions \cite{Jaubert09, Jaubert11} at the lowest temperatures. 

In the QO-FeA, the presence of the fluctuating boundary spins is further evidenced in the magnetization measurements. Below 2 K, a small field-induced magnetization ($<$0.5 $\mu_B$) appears in the foot of the magnetization curves (see Fig.  \ref{figMH}(b)). It is compatible with $\approx$ 5 \% of field polarized quasi-Ising free spins as shown in the calculation of the inset of Fig. \ref{figMH}(a). This field-induced magnetization explains the upturn of the susceptibility below 2 K before it reaches a second maximum at 400 mK (See Fig. \ref{figXT}). Below this temperature, a second metamagnetic transition is observed at a small field of 0.1 T (see Fig. \ref{figMH}(b)), consistent with the dipolar energy scale. The onset of dipolar interactions does not yield such a metamagnetism in the magnetization curves computed for the ordered spins inside the domains. This rather originates from the field-response of the dipolar coupled quasi-Ising spins along the domain walls. 

These new QO-FeA compounds have enabled the investigation of a magnetic (distorted) kagome lattice with exchange and dipolar interactions, and multiaxial anisotropy, the same ingredients as in the pyrochlore spin ices. At variance with those, a non-collinear magnetic ordering is favored. The water/magnetic ice analogy can nevertheless be pursued on a larger nanometric scale with sea ice: in a porous solid ice matrix, liquid inclusions flow in the interstices, and get amorphously frozen as the temperature is decreased  \cite{mushy}. The topology of the kagome lattice provides a medium similarly sustaining free spins at the boundary between the antiferromagnetic domains, before they become correlated through dipolar interactions. An assembly of magnetic nanocrystals, related to each other by the time-reversal symmetry operation, but magnetically decoupled due to interstitial paramagnetic spins, is thus achieved, providing an example of a classical protectorate of free spins \cite{protectorate}. The observed dynamics could be generic to geometrically frustrated magnets, where residual spin fluctuations often are observed in the ordered phase \cite{Mirebeau}. These results could also be relevant for applications utilizing frustrated magnets like multiferroics where the magnetoelectric manipulation of magnetic/ferroelectric domains is foreseen \cite{multiferroics}. 


\acknowledgments We would like to thank D. Givord, P. Molho, J-.J. Pr\'ejean and C. Train for fruitful discussions, and P. Convert for his help during the first neutron experiment on D20. DJP, PTW and AKP are grateful for financial support from the EPSRC.


\begin{thebibliography}{}
\bibitem{footnote2} Within this assumption, the $J_2$ exchange interaction is absorbed in an effective NN exchange interaction.
\bibitem{footnote3} This is true of the $J_1$ interaction, but also of the weaker $J_2$ interaction when the domain wall makes up a straight line in the kagome plane. $J_2$ is relevant only at wall corners.
\end{thebibliography}
\end{document}